\documentclass[%
  twoside,
  reprint,
  amsmath,amssymb,
  aps,
  pra,
  nofootinbib,
  showpacs,
  superscriptaddress,
  a4paper
]{revtex4-1}

\usepackage{graphicx}%
\usepackage[usenames,dvipsnames]{xcolor}
\usepackage{siunitx}
\usepackage{subfigure}

\usepackage{titlesec}

\usepackage{array}

\usepackage[utf8]{inputenc}
\usepackage[T1]{fontenc}

\makeatletter
\renewcommand\frontmatter@abstractwidth{\dimexpr\textwidth-1.2in\relax}
\makeatother

\usepackage{lipsum}

\graphicspath{{Figures/}{}}

\usepackage[
centering, includefoot,
text={7.1in,10.2in},
total={6.3in,8.75in},
top=0.8in, left=0.62in,
]{geometry}

\usepackage[
  bookmarks=false,
  colorlinks,
  linkcolor=blue,
  urlcolor=blue,
  citecolor=blue,
  plainpages=false,
  pdfpagelabels,
  final,
  breaklinks=true
]{hyperref}
\hypersetup{
pdftitle={Conservation of torus-knot angular momentum in high-order harmonic generation}, 
pdfauthor={Emilio Pisanty, Laura Rego, Julio San Román, Antonio Picón, Kevin M. Dorney, Henry C. Kapteyn, Margaret M. Murnane, Luis Plaja, Maciej Lewenstein and Carlos Hernández-García}
}

\usepackage{natbib}
\makeatletter \def\NAT@def@citea{\def\@citea{\NAT@separator\,}} \makeatother
\newcommand{\citer}[1]{Ref.~\citealp{#1}}
\newcommand{\citers}[1]{Refs.~\citealp{#1}}

\newcommand{\reffig}[1]{Fig.~\ref{#1}}

\renewcommand{\d}{\ensuremath{\textrm{d}}}
\renewcommand{\Re}{\operatorname{Re}}

\usepackage{physics}
  \newcommand{\vbr}{\vb{r}}

  \newcommand{\vbf}{\vb{F}}

\newcommand{\ue}[1]{\hat{\vb{e}}_{#1}}

\newcommand{\Ri}{R^{-1}\mspace{-2mu}}

\newcommand{\xuv}{\mathsf{(xuv)}}

\DeclareFontFamily{U} {MnSymbolA}{}
\DeclareFontShape{U}{MnSymbolA}{m}{n}{
  <-6> MnSymbolA5
  <6-7> MnSymbolA6
  <7-8> MnSymbolA7
  <8-9> MnSymbolA8
  <9-10> MnSymbolA9
  <10-12> MnSymbolA10
  <12-> MnSymbolA12}{}
\DeclareSymbolFont{MnSyA} {U} {MnSymbolA}{m}{n}
\DeclareMathSymbol{\rcirclearrowleft}{\mathrel}{MnSyA}{250}
\DeclareMathSymbol{\lcirclearrowright}{\mathrel}{MnSyA}{252}

\newcommand{\leftpol}{\!\; \!\!\lcirclearrowright \!}
\newcommand{\rightpol}{\!\; \!\!\rcirclearrowleft \!}

\begin{document}

\title{Conservation of torus-knot angular momentum in high-order harmonic generation
}

\author{Emilio Pisanty}
 \affiliation{ICFO -- Institut de Ciencies Fotoniques, The Barcelona Institute of Science and Technology, 08860 Castelldefels (Barcelona)}

\author{Laura Rego}
 \affiliation{Grupo de Investigación en Aplicaciones del Láser y
Fotónica, Departamento de Física Aplicada, University of Salamanca, E-37008, Salamanca, Spain.}

\author{Julio San Román}
 \affiliation{Grupo de Investigación en Aplicaciones del Láser y
Fotónica, Departamento de Física Aplicada, University of Salamanca, E-37008, Salamanca, Spain.}

\author{Antonio Picón}
 \affiliation{ICFO -- Institut de Ciencies Fotoniques, The Barcelona Institute of Science and Technology, 08860 Castelldefels (Barcelona)}
 \affiliation{Departamento de Química, Universidad Autónoma de Madrid, 28049, Madrid, Spain}

\author{Kevin M. Dorney}
 \affiliation{JILA, Department of Physics, University of Colorado Boulder, Boulder, Colorado, 80309, USA}

\author{Henry C. Kapteyn}
 \affiliation{JILA, Department of Physics, University of Colorado Boulder, Boulder, Colorado, 80309, USA}

\author{Margaret M. Murnane}
 \affiliation{JILA, Department of Physics, University of Colorado Boulder, Boulder, Colorado, 80309, USA}

\author{Luis Plaja}
 \affiliation{Grupo de Investigación en Aplicaciones del Láser y
Fotónica, Departamento de Física Aplicada, University of Salamanca, E-37008, Salamanca, Spain.}

\author{Maciej Lewenstein}
 \affiliation{ICFO -- Institut de Ciencies Fotoniques, The Barcelona Institute of Science and Technology, 08860 Castelldefels (Barcelona)}
 \affiliation{ICREA, Passeig de Lluís Companys, 23, 08010 Barcelona, Spain}

\author{Carlos Hernández-García}
 \affiliation{Grupo de Investigación en Aplicaciones del Láser y
Fotónica, Departamento de Física Aplicada, University of Salamanca, E-37008, Salamanca, Spain.}

\date{10 June 2019}

\begin{abstract}
High-order harmonic generation stands as a unique nonlinear optical up-conversion process, mediated by a laser-driven electron recollision mechanism, which has been shown to conserve energy, momentum, and spin and orbital angular momentum.
Here we present theoretical simulations which demonstrate that this process also con\-ser\-ves a mixture of the latter, the torus-knot angular momentum $J_\gamma$, by producing high-order harmonics with driving pulses that are invariant under coordinated rotations. 
We demonstrate that the charge $J_\gamma$ of the emitted harmonics scales linearly with the harmonic order, and that this conservation law is imprinted onto the polarization distribution of the emitted spiral of attosecond pulses. 
We also demonstrate how the nonperturbative physics of high-order harmonic generation affect the torus-knot angular momentum of the harmonics,
and we show that this configuration harnesses the spin selection rules to channel the full yield of each harmonic into a single mode of controllable orbital angular momentum.
\\[-2mm]

\noindent
\footnotesize
Accepted manuscript for
\href{%
  https://doi.org/10.1103/PhysRevLett.122.203201%
  }{%
  \color[rgb]{0,0,0.55}%
  \textit{Phys.\ Rev.\ Lett.} \textbf{122}, 203201  (2019)%
  }. 
Available as %
\href{%
  https://arxiv.org/abs/1810.06503%
  }{%
  \color[rgb]{0,0,0.55}%
  arXiv:1810.06503%
  } %
under %
\href{%
  %
  https://creativecommons.org/licenses/by-nc-sa/4.0/
  }{%
  \color[rgb]{0,0,0.55}%
  CC BY-NC-SA%
  }.
\\[-6mm]

\end{abstract}

\maketitle

Nonlinear optical processes offer the unique possibility of mediating interactions and transferring energy between modes of the electro\-magnetic field at different frequencies~\cite{boyd-nonlinear-optics-2003}. When this transfer happens in a symmetric medium, the interaction will also carry the symmetry's conserved charge to the recipient mode~\cite{neuenschwander-noether-2011, bloembergen-conservation-1980}, so one can e.g.\ combine two photons with well-defined orbital angular momentum (OAM)~\cite{molina-twisted-photons-2007} to make a single photon at twice the frequency and twice the angular momentum~\cite{courtial-shg-oam-1997}.
The few-photon exchanges of perturbative nonlinear optics, however, have a relatively limited scale and complexity in comparison to high-order harmonic generation (HHG) \mbox{[\citealp{ivanov-tutorial-2014, krausz-ivanov-attosecond-review-2009}]}, where strong-field interactions can produce harmonics with photon energy hundreds or thousands of times larger than the driver~\cite{popmintchev_record_2012}.
HHG is a nonperturbative phenomenon which is best understood using a semiclassical picture: an ionized electron is accelerated back to its parent ion, emitting high-frequency light in the ensuing recollision~\cite{lewenstein-hhg-1994, corkum-plasma-perspective-1993, schafer-ati-1993}.
Despite the lack of a photon-exchange model, HHG is often regarded as a parametric process, and its conservation properties have been explored extensively as regards energy~\cite{perry-hhg-energy-conservation-1993}, linear momentum~\cite{bertrand-hhg-momentum-conservation-2011}, and orbital and spin angular momentum (SAM)~[\citealp{zurch-hhg-oam-conservation-2012, hernandez-attosecond-vortices-2013, gariepy-creating-2014, rego-nonperturbative-twist-2016, geneaux-synthesis-2016, kong-controlling-2017, gauthier-tunable-2017, hernandez-hhg-twist-2017},\citealp{fleischer-spin-2014, ivanov-taking-control-2014, pisanty-spin-conservation-2014, hickstein-non-collinear-2015, huang-polarization-control-2018}].

The individual symmetries associated with these conservation laws of the electromagnetic field can be composed in nontrivial ways to make new ones. This is the case for \textit{coordinated} rotations~(CRs): symmetry transformations in which the spatial dependence of the field is rotated by an angle~$\theta$ about the propagation axis, while the light's polarization is rotated by $\gamma\mspace{1.5mu}\theta$ around the same axis, $\gamma$ being a coordination parameter. CRs are generated by the linear combination $J_\gamma = L+\gamma \mspace{1.5mu} S$ of the orbital and spin angular momenta, $L$ and~$S$, which are otherwise independently conserved in the paraxial regime~\cite{leach-interferometric-2004}. 
For mono\-chromatic light, $\gamma$ is restricted to integer or half-integer values~\cite{ballantine-many-ways-to-spin-a-photon-2016, dennis-morphology-2002}, with the latter case imparting on the field the topology of a Möbius strip~\cite{freund-classification-cones-spirals-strips-2005, bauer-observation-mobius-2015, bauer-optical-mobius-2016}.
However, when the mono\-chro\-matic restriction is lifted, $\gamma$ can take arbitrary values and still admit invariant states of the field~\cite{pisanty-knots-2019}, since polychromatic combinations can have polarization states with higher-order internal rotational symmetries.

One particularly relevant example is the three-fold-symmetric trefoil field present in the `bicircular' HHG configurations~\cite{fleischer-spin-2014, ivanov-taking-control-2014, pisanty-spin-conservation-2014, hickstein-non-collinear-2015, huang-polarization-control-2018, eichmann-polarization-1995, milosevic-apts-unusual-polarization-2000, kfir-generation-2015, fan-bright-circularly-2015, chen-tomographic-2016, pisanty-rotating-frame-2017, jimenez-control-2018, dorney-controlling-2019} used to produce circularly-polarized harmonics. This field consists of two counter-rotating circularly-polarized drivers at different frequencies, and exhibits the same configuration after a polarization rotation by an angle $2\pi/n$, with $n\geq 3$. If we then add different OAM to the two drivers, this polarization rotation can be realized over a $2\pi$ rotation of the spatial dependence. 
The resulting field, which carries both SAM and OAM, but is not an eigenstate of either, has the topology of a torus knot: when the polarization and spatial dependences of the field are unfolded, the trefoil tips trace out a knotted curve embedded on the surface of a torus~\cite{pisanty-knots-2019}. With suitable choices of the OAM and frequency of the two components, any arbitrary torus knot~\cite{adams-knot-book-2004} can be achieved. 
Moreover, this topology mirrors the subgroup of the independent-rotations group $\mathrm{SO}(2)\times\mathrm{SO}(2)$ generated by the torus-knot angular momentum (TKAM) $J_\gamma = L+\gamma \mspace{1.5mu} S$~\cite{pisanty-knots-2019}.

\begin{figure*}[ht]
\setlength{\tabcolsep}{0mm}
\begin{tabular}[b]{m{37mm}m{138mm}}
\begin{tabular}[b]{c}
\includegraphics[scale=1]{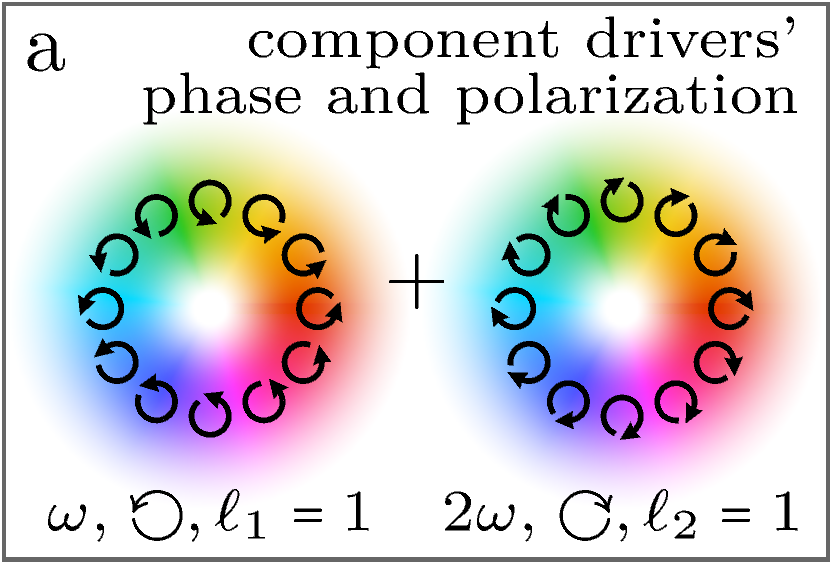} \\
\includegraphics[scale=1]{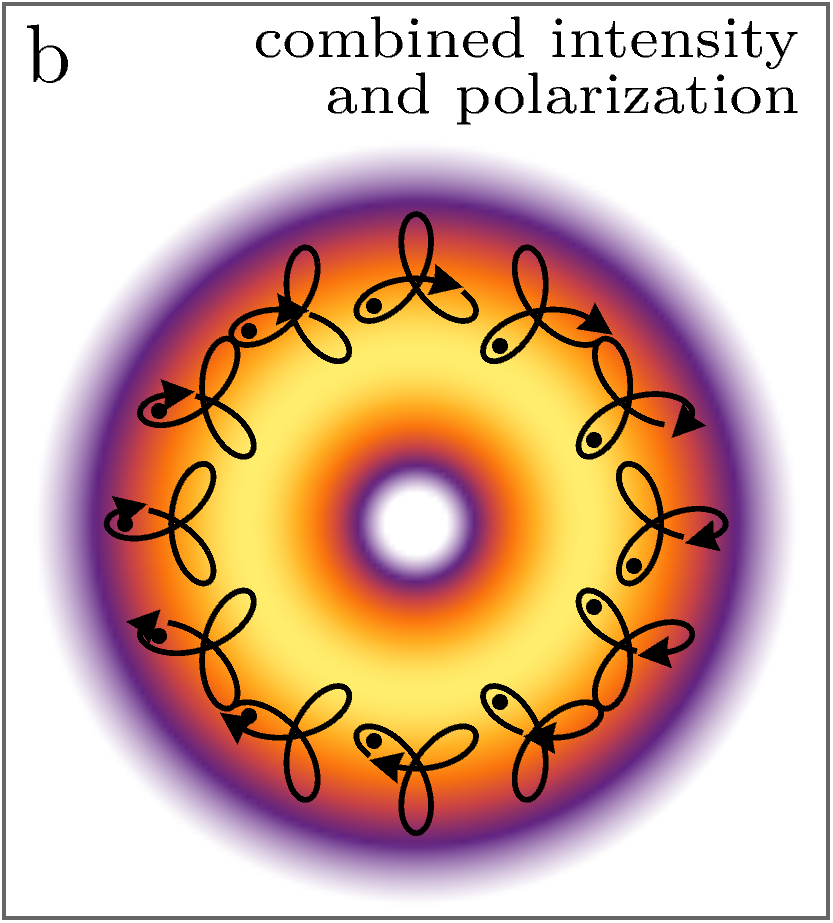}
\end{tabular} &
\includegraphics[scale=1]{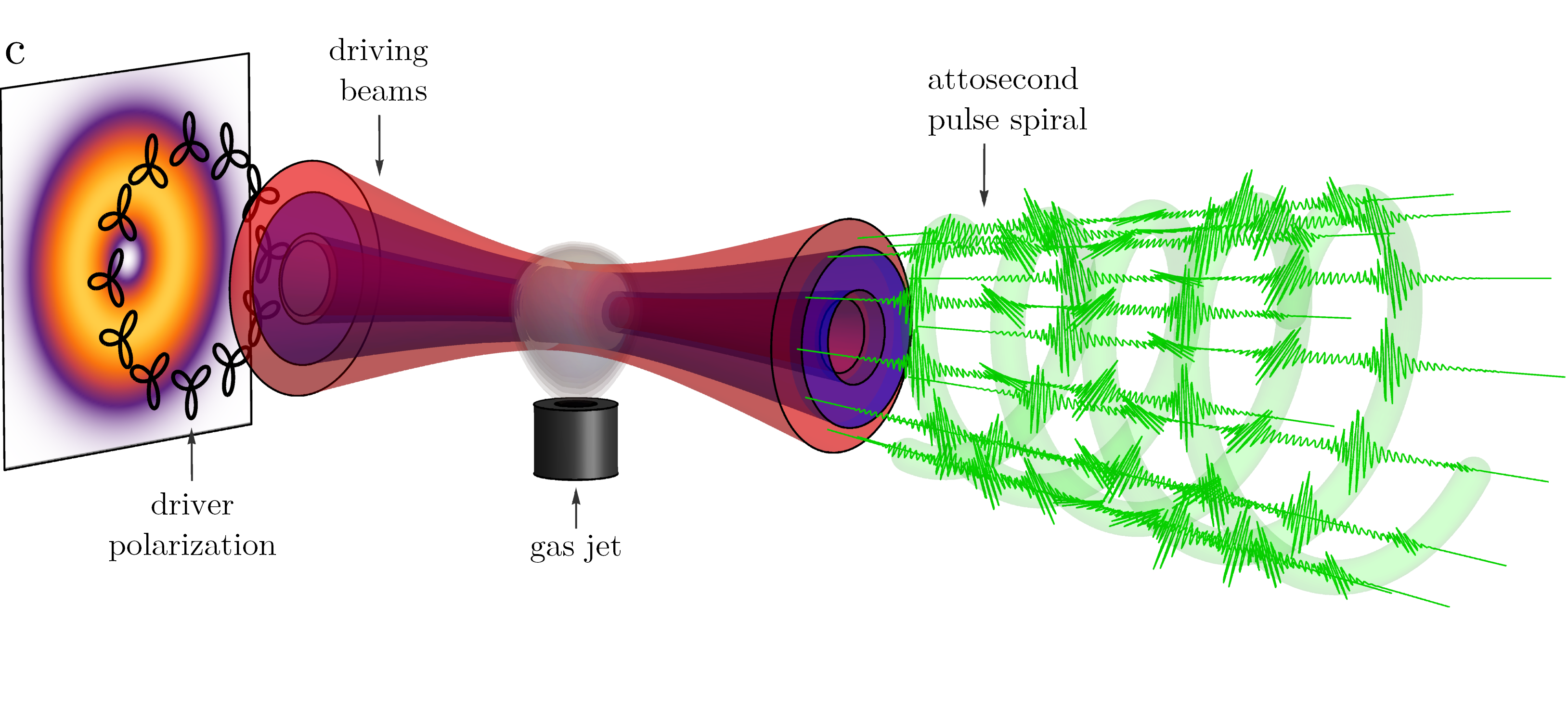}
\vspace{-10mm}
\end{tabular}
\caption{
High-harmonic generation driven by a torus-knot beam. The combination of counter-rotating circular beams at frequencies $\omega$ and $2\omega$ with different OAM and therefore different azimuthal phase gradients ((a), with optical phase as the hue color scale) produces a bicircular trefoil polarization which rotates and acquires a delay over azimuthal displacements (b). Here, with $\ell_1=\ell_2=1$, tracking one lobe over an azimuthal loop around the beam axis (black dots) produces a $\SI{120}{\degree}$ rotation, which induces the topology of a torus knot~\cite{pisanty-knots-2019}, as well as a time delay within each trefoil (arrows).  When this combination incides on a gas jet at high intensity (c), the attosecond pulse trains produced share the coordinated-rotation invariance of the driver, so that the emission at different azimuthal points is related by a time delay and a rotation of the polarization.
}
\label{fig-configuration}
\end{figure*}

In this work, we show that the topological charge $J_\gamma$ is conserved in high-order harmonic generation for any arbitrary rational coordination parameter $\gamma\in\mathbb{Q}$~\cite{Note1}, 
providing an infinite family of topological charges (corresponding to an infinite family of dynamical symmetries) that are preserved by the nonperturbative nonlinear interaction. 
We show that this spin-orbit linking appears in the time domain as a structured spiral of attosecond pulses, that it opens a new window for the exploration of nonperturbative effects in harmonic generation, and that it provides an XUV light source with controllable OAM.

To demonstrate this conservation property, we study HHG driven by TKAM beams that are invariant under coordinated rotations with a mixing parameter $\gamma\in \frac13\mathbb Z$. This configuration corresponds, as we will show below, to a bicircular field consisting of two beams at frequencies $\omega$ and $2\omega$, with counter-rotating right- and left-handed circular polarizations, $\rightpol$ and $\leftpol$, and carrying independent OAM, $\ell_1$~and~$\ell_2$, respectively, as shown in \reffig{fig-configuration}. The OAM of the drivers determines the mixing parameter $\gamma$ under which the bicircular beam is CR-invariant, so that the two components carry a TKAM of $j_{\gamma}^{(1)}=\ell_1 + \gamma$ and $j_{\gamma}^{(2)}=2j_{\gamma}^{(1)} = \ell_2 -\gamma$, respectively~\cite{Note4}. 
(The drivers are required to carry different TKAM to have identical CR dynamical symmetries, since phase and time delays correspond differently at different frequencies.)
Within that framework, then, the TKAM conservation is expressed~as
\begin{equation}
j_{\gamma}^{(q)} = q \, j_{\gamma}^{(1)},
\label{tkam-conservation}
\end{equation}
i.e.\ in the linear scaling of the $J_\gamma$ charge carried by the $q^\mathrm{th}$ harmonic with the harmonic order $q$.

In this configuration, the local field at each point in the beam is the usual bicircular trefoil~\cite{fleischer-spin-2014}, so that each atom in the target emits harmonics in circularly-polarized doublets with opposite helicities: $\rightpol$-polarized harmonics at frequencies $(3n+1)\omega$, and $\leftpol$-polarized harmonics at $(3n-1)\omega$; in the time domain, the emission forms a train of attosecond pulses with linear polarizations at $\SI{120}{\degree}$ from each other~\cite{ milosevic-apts-unusual-polarization-2000, chen-tomographic-2016}. The orientation of this local trefoil, given by the relative phase between the two components rotates around the beam, as shown in \reffig{fig-configuration}(b).

Thus, the dynamical symmetry of the driving field~$\vbf$~is
\begin{equation}
R(\gamma \alpha)\vbf \big(\Ri(\alpha)\vbr,t \big)
=
\vbf(\vbr, t+\tau \alpha)
,
\label{core-invariance-equation}
\end{equation}
where $\tau$ is a time-delay constant and the angle $\alpha$ parametrizes the transformation. Here the rotations act on the circular polarization basis $\ue\pm = \frac{1}{\sqrt{2}}(\ue{x} \pm i \ue{y})$ and on the spatial dependence via
\begin{subequations}%
\begin{align}%
R(\gamma\alpha) \ue\pm
& =
e^{\mp i \gamma\alpha} \ue\pm
\quad \text{and}
\\
\Ri(\alpha) (r,\theta, z)
& = 
(r,\theta-\alpha,z),
\end{align}%
\end{subequations}%
%
%
%
%
i.e.\ as an active and a passive transformation, respectively, with the polarization rotation $R(\gamma \alpha)$ acting through a fraction $\gamma \alpha$ of the spatial rotation angle $\alpha$.

Our driving field consists of two components with well-defined SAM and OAM,
\begin{subequations}%
\begin{align}%
\vbf_1(\vbr,t) & = \Re\left[F_1 \ue{+} f_1(r,z) e^{i\ell_1\theta} e^{-i\omega t}\right] \\
\vbf_2(\vbr,t) & = \Re\left[F_2 \ue{-} f_2(r,z) e^{i\ell_2\theta} e^{-2i\omega t}\right],
\end{align}%
\label{field-components-definition}%
\end{subequations}%
each of which satisfies separate orbital and spin invariance properties
\begin{subequations}%
\begin{align}
R(\gamma \alpha)\vbf_1(\vbr,t) 
\phantom{\Ri(\alpha)}
& = 
\vbf_1(\vbr,t + \gamma\alpha/\omega) 
,
\label{individual-action-spin-f1}
\\
\vbf_1(\Ri(\alpha)\vbr,t) 
& = 
\vbf_1(\vbr,t + \ell_1\alpha/\omega)
,
\label{individual-action-oam-f1}
\\
R(\gamma \alpha)\vbf_2(\vbr,t) 
\phantom{\Ri(\alpha)}
& = 
\vbf_2(\vbr,t - \gamma\alpha/2\omega)
,
\label{individual-action-spin-f2}
\\
\vbf_2(\Ri(\alpha)\vbr,t) 
& = 
\vbf_2(\vbr,t + \ell_2\alpha/2\omega)
.
\label{individual-action-oam-f2}%
\end{align}%
\label{individual-actions}%
\end{subequations}%
\vspace{-12pt}

The correct CR invariance of the system can then be found by requiring that the combined time delay imposed by Eqs.~(\ref{individual-action-spin-f1},\,\ref{individual-action-oam-f1}) matches that produced by the combination of Eqs.~(\ref{individual-action-spin-f2},\,\ref{individual-action-oam-f2}), so that
\begin{align}
\frac{\ell_1\alpha}{\omega} + \frac{\gamma\alpha}{\omega}
& =
\frac{\ell_2\alpha}{2\omega} - \frac{\gamma\alpha}{2\omega}
\ \implies \ 
\gamma
 =
\frac{\ell_2 - 2\ell_1}{3},
\end{align}
with a time-delay constant $\tau = \frac{\ell_1 + \ell_2}{3\omega}$. This then sets the TKAM charge $j_{\gamma}^{(n)}$ for each driver, defined in analogy to the OAM charge in Eqs.~(\ref{individual-action-oam-f1},\,\ref{individual-action-oam-f2}) by requiring that 
\begin{equation}
R(\gamma \alpha)\vbf_n \big(\Ri(\alpha)\vbr,t \big)
=
\vbf(\vbr, t+ j_{\gamma}^{(n)} \alpha/n\omega ),
\label{cr-invariance-with-tkam-charge}
\end{equation}
to be $j_{\gamma}^{(n)} = n \omega \tau = n\tfrac{\ell_1 + \ell_2}{3}$~\cite{Note2}.%

\begin{figure}[t!]
\begin{center}
\begin{tabular}{cc}
\includegraphics[scale=1]{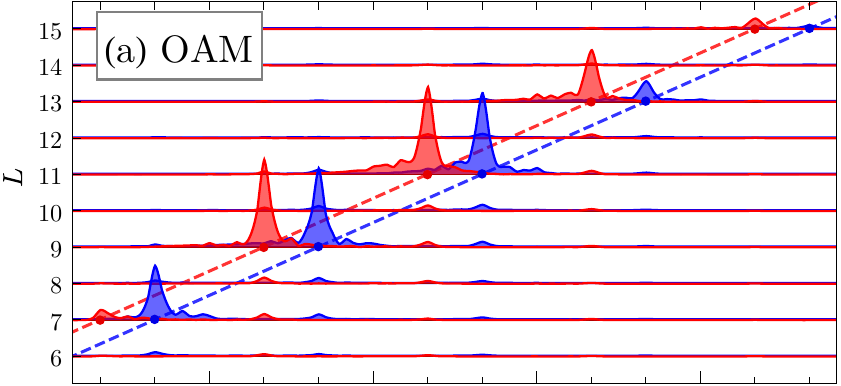} \\
\includegraphics[scale=1]{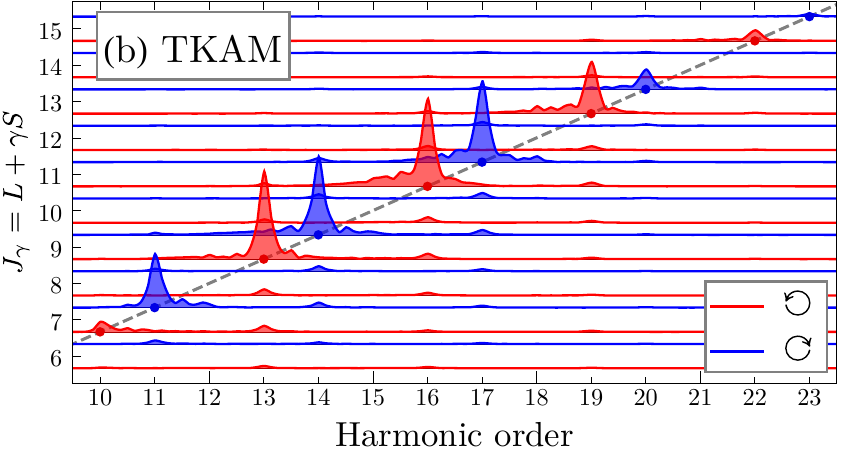}
\end{tabular}
\vspace{-6mm}
\end{center}
\caption{%
Simulated HHG spectra of the right and left-polarized components (shown in red and blue, respectively) driven by a bicircular field with $\ell_1=\ell_2=1$. We compare the (a) OAM and (b)~TKAM spectra, with $\gamma=-1/3$, as a function of the harmonic order. We calculate the OAM spectrum by taking a standard Fourier series over the azimuthal dependence and integrating over the radial dependence. For the TKAM spectrum the OAM charge of the $\rightpol$- and $\leftpol$-polarized components is shifted by $\gamma$ times their SAM. The conservation of TKAM is clear in the linear trend shown in~(b).
}
\vspace{-1em}
\label{fig-results-charge-vs-harmonic-order}
\end{figure}

Turning to the HHG radiation, we can now see the conserved TKAM charge in action, via the standard correspondence between dynamical symmetries and conserved charges, and their associated selection rules~\cite{alon-selection-rules-1998, averbukh-stability-2002}. Since in our configuration the application of a CR to the driving field is equivalent to a time delay, via Eq.~\eqref{core-invariance-equation}, and the gaseous generating medium is unaffected by the transformation, the same must be true for the emitted HHG radiation. With the XUV emission's TKAM charge defined as in Eq.~\eqref{cr-invariance-with-tkam-charge}, the CR invariance then guarantees the conservation of the TKAM charge as expressed in~Eq.~\eqref{tkam-conservation}.

To explore the conservation of TKAM in HHG, we perform numerical simulations of HHG by solving the Schr\"odinger-Maxwell equations for a sample of atoms in the interaction region within the SFA\texttt{+} approximation and using the electromagnetic field propagator described in \citer{hernandez-discrete-dipole-2010}; further details of our method can be found in \citers{hernandez-attosecond-vortices-2013, rego-nonperturbative-twist-2016, hernandez-quantum-path-2015}.

We consider the harmonic emission driven by a bicircular field with $\ell_1=\ell_2=1$, as shown in \reffig{fig-configuration}, equal beam waists of $\SI{30}{\micro m}$, and pulses of total intensity $I=\SI{2e14}{W/cm^2}$.
The $\omega$ and $2\omega$ driving pulse envelopes are modelled as a trapezoidal function with $\SI{5.3}{fs}$ linear on- and off-ramps and $\SI{10.7}{fs}$ of constant amplitude.
Harmonics are generated in a thin-slab argon gas jet and propagated to the far-field (i.e.\ longitudinal phase-matching effects are neglected, since transverse phase-matching effects are dominant~\cite{rego-nonperturbative-twist-2016,hernandez-quantum-path-2015}).

\begin{figure}[t!]
\begin{tabular}{l}
\includegraphics[width=0.47\textwidth]{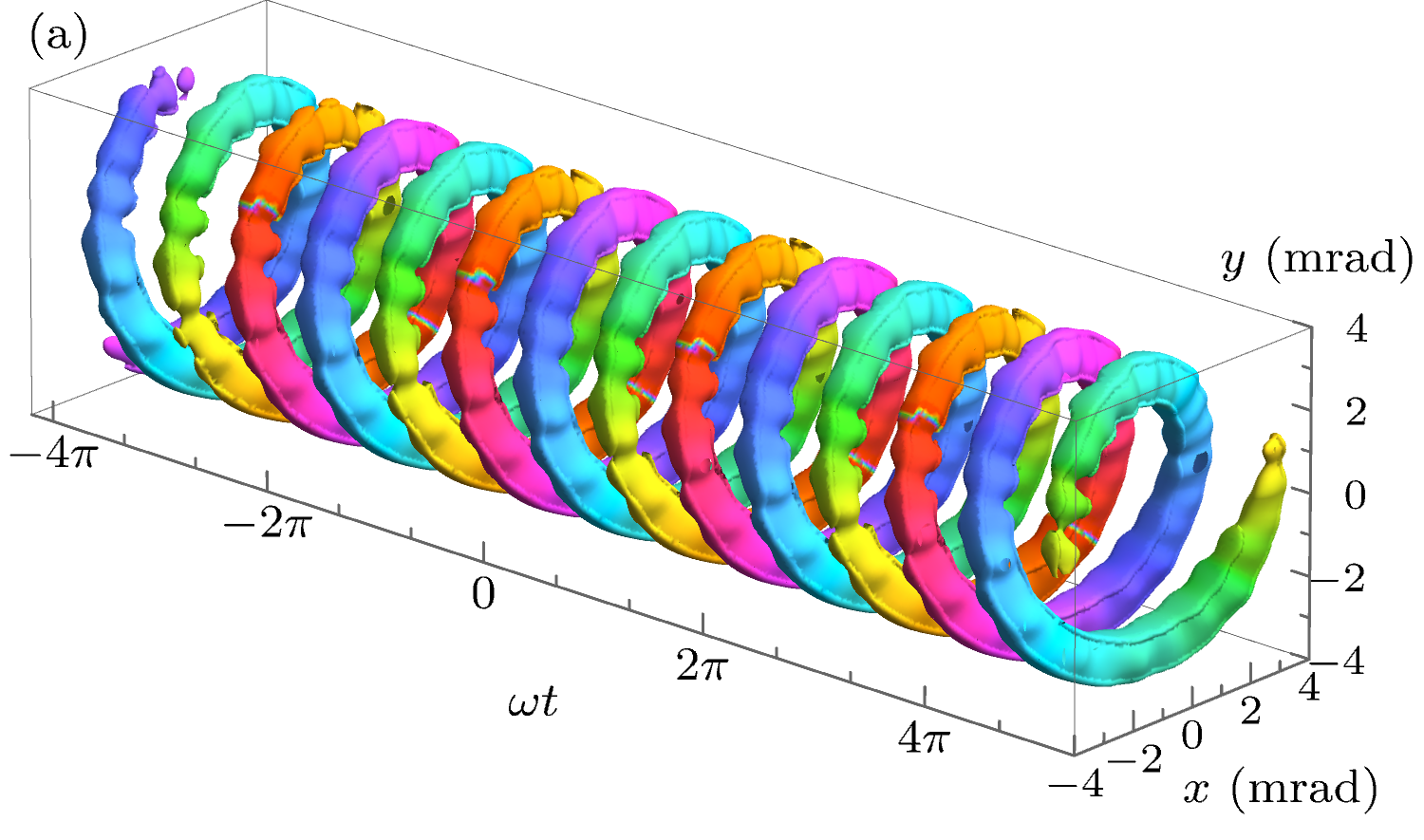}
\hspace{-0.47\textwidth}\raisebox{3mm}{%
\includegraphics[scale=1]{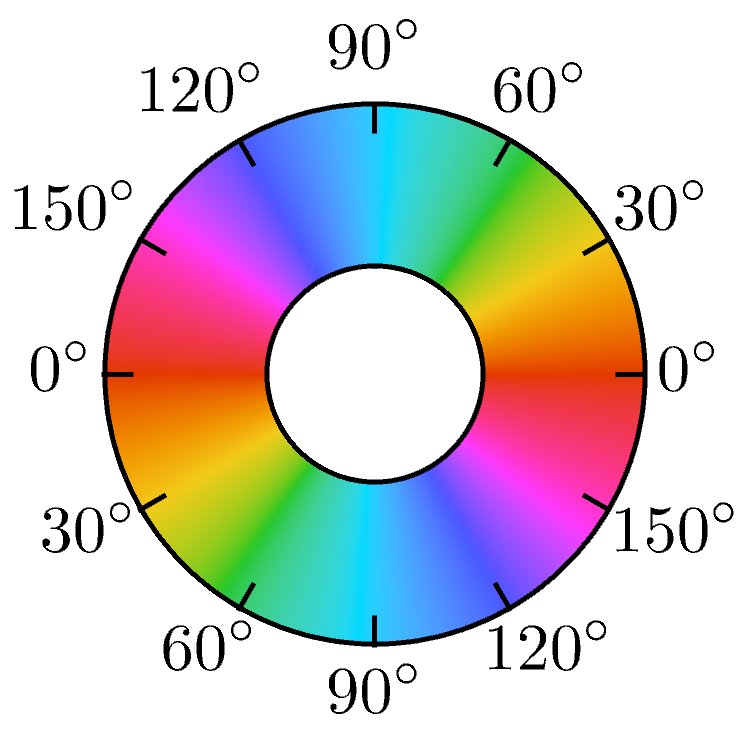}%
}%
\\
\includegraphics[scale=1]{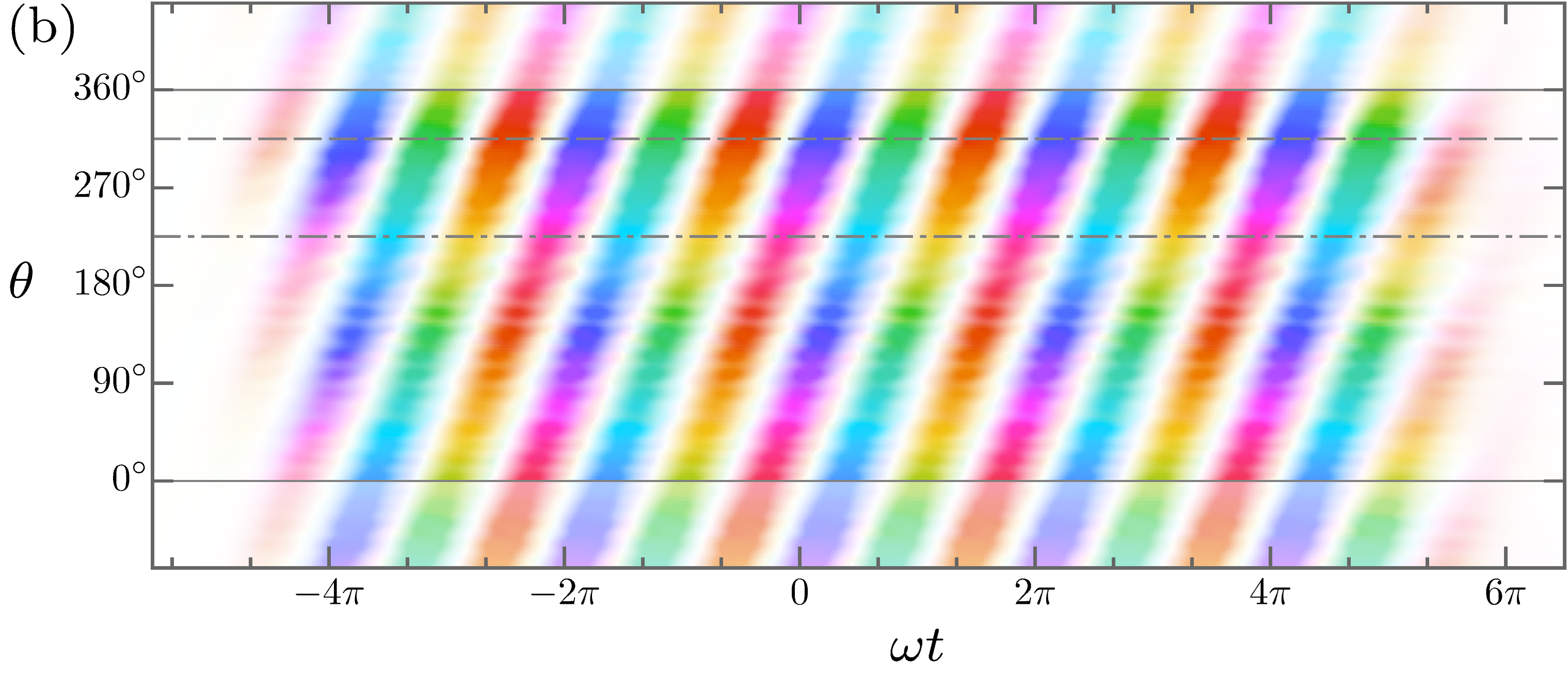}
\end{tabular}
\caption{%
Twisted-spiral structure of the attosecond pulse train emitted by CR-invariant bicircular drivers at $\ell_1=\ell_2=1$, as in \reffig{fig-results-charge-vs-harmonic-order}. 
(a) Iso-intensity surface of the HHG emission, filtered above harmonic order 10, over the far-field divergence Cartesian coordinates $x$ and $y$, with the color denoting the local polarization direction as in (b).
(b) Polarization angle of the attosecond pulses over azimuthal emission angle $\theta$, obtained from the time-windowed $T_{22}(\vbr,t)$ field moment of Eq.~\eqref{time-windowed-t22} with $\sigma=\SI{15}{\degree}/\omega$ and integrated over angular divergence, with $\frac12\arg(T_{22}(\vbr,t))$ plotted as the hue and $|T_{22}(\vbr,t)|$ (which closely follows the XUV intensity) as the color saturation.
The polarization angle rotates by $-\SI{120}{\degree}$ over each turn of the spiral, directly confirming the $\gamma=-1/3$ rotation-coordination parameter of the beam.
At each fixed $\theta$, the polarization jumps over three complementary colors such as blue-green-red (dashed line) or cyan-yellow-magenta (dot-dashed line), showing the local polarization structure as a train of linear pulses at $\SI{120}{\degree}$ from each other.
}
\label{fig-spiral-structure}
\end{figure}

We present our results in \reffig{fig-results-charge-vs-harmonic-order}, by comparing the (a) OAM and (b) TKAM spectra of the circularly polarized components of the HHG emission as a function of harmonic order. 
Since in this configuration $j_\gamma^{(1)}=2/3$, following Eq.~\eqref{tkam-conservation} the $q^\mathrm{th}$ harmonic exhibits a TKAM of $j_\gamma^{(q)}= \tfrac{2}{3} q$ (so e.g.\ $j_\gamma^{(13)}= \frac{26}{3}$ and $j_\gamma^{(14)}= \tfrac{28}{3}$) and, following the definition of the TKAM, its OAM would be $\ell_q = j_\gamma^{(q)} - \gamma S_q = (2q\pm 1)/3$ for right- ($+$) and left- ($-$) polarized harmonics (therefore giving $\ell_{13} = \ell_{14} = 9$, in agreement with the results of Refs.~\citealp{dorney-controlling-2019, paufler-oam-hhg-2018}).
Thus, while the OAM behaviour can be explained using photon-counting methods, the spectrum is much easier to understand via the TKAM conservation law, which is embodied in the linear trend observed in the TKAM spectrum, as in Eq.~\eqref{tkam-conservation}. 
A similar conservation law, with $\gamma=1/2$, can also be observed in HHG driven by monochromatic CR-invariant beams~\cite{turpin-xuv-fractional-oam-2017}.

The CR invariance can also be seen in the time domain: similarly to OAM-beam HHG, where a spatial rotations are equivalent to time delays and the attosecond pulse train (APT) forms a spiral~\cite{hernandez-attosecond-vortices-2013, hernandez-quantum-path-2015, geneaux-synthesis-2016},
the addition of a polarization rotation means that here the spiral also has a twisted-polarization structure~\cite{turpin-xuv-fractional-oam-2017}.
We present this in \reffig{fig-spiral-structure}(a), using an isosurface plot for the XUV intensity and the color scale to represent the polarization direction, which twists smoothly along the attosecond pulse spiral.

To study this polarization structure quantitatively, we require a measure of the absolute orientation of the APT at different azimuthal points in the beam---and, ideally, one which is sensitive to the pulse train's structure as a sequence of linearly-polarized pulses at nontrivial angles~\cite{chen-tomographic-2016, milosevic-apts-unusual-polarization-2000}.
For monochromatic radiation, the polarization ellipse orientation angle is obtained via the eigenvectors of the polarization matrix $\langle E_iE_j\rangle$~\cite{freund-coherency-matrix-2004} or, equivalently, as the phase of the quadrupole component $T_{22} = \int_{-\infty}^\infty(E_x(t) + i E_y(t))^2\d t$~\cite{pisanty-knots-2019}. The symmetry of our APT means that the average orientation is undefined, so we use a time-windowed version~\cite{antoine-time-profile-1995},
\begin{align}
\label{time-windowed-t22}
T_{22}(\vbr, t)
& =
\int_{-\infty}^\infty
(E_x^\xuv(\vbr, t') + i E_y^\xuv(\vbr, t') )^2
\\ \nonumber & \qquad \qquad \qquad \qquad \qquad \times
e^{-(t'-t)^2/2\sigma^2}
\d t'
,
\end{align}
from which the local polarization orientation angle can be obtained as $\tfrac12 \arg(T_{22}(\vbr, t))$~\cite{Note3}. %
Thus, for example, a linearly polarized pulse along $\ue{x}$ produces a positive $T_{22}$, an $\ue{y}$ polarization gives a negative moment, and there is a continuous passage between the two behaviours.

The time-windowed quadrupole moment, which we plot in \reffig{fig-spiral-structure}(b), clearly shows the twisted-spiral structure of the XUV emission: the amplitude is constrained to two strips that wind around the azimuthal axis, acquiring a time delay of $4\pi/3\omega$ after one revolution, while the polarization direction, indicated by the hue color scale, turns by $-\SI{120}{\degree}$ over that span, in an essentially linear progression. This directly confirms the CR invariance of the HHG emission, in the sense of Eq.~\eqref{core-invariance-equation}, and, with that, its nontrivial torus-knot topology.

\begin{figure}[t]
\begin{tabular}{c}
\includegraphics[scale=1]{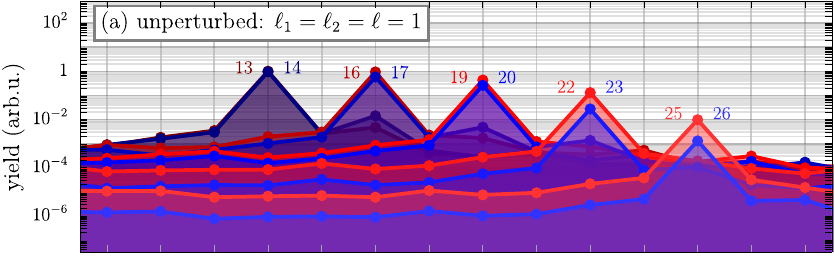} \\[-1mm]
\includegraphics[scale=1]{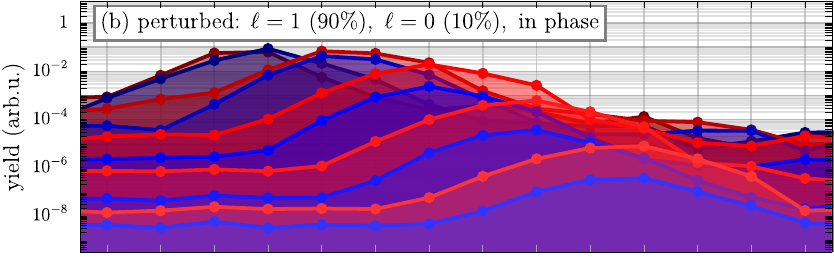} \\[-1mm]
\includegraphics[scale=1]{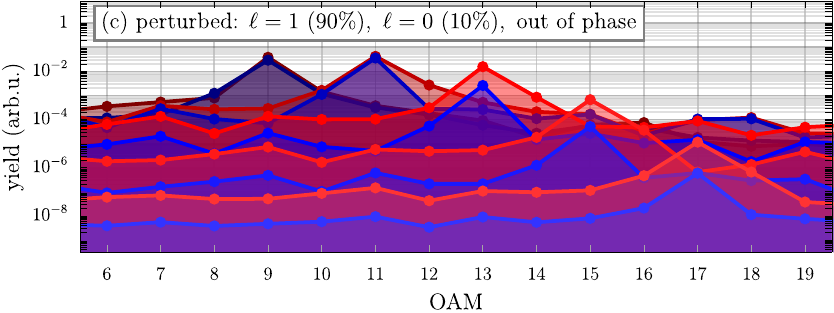}
\end{tabular}
\caption{%
OAM spectrum of the HHG radiation driven by a bicircular field with $\ell_1=\ell_2=1$ (a), as in \reffig{fig-results-charge-vs-harmonic-order}, as well as with a $10\%$ intensity perturbation on a donut-like $\ell=0$ mode on both drivers: (b) with both perturbations in phase, with matching intensity profiles, and (c) with the perturbations in opposite phase, with complementary intensity profiles.
(The lines are color-coded by harmonic order, as in (a).)
}
\label{fig-perturbed-symmetries}
\end{figure}

Here it is also instructive to broaden our scope to consider what happens when the coordinated-rotation invariance gets broken, by perturbing the OAM of one or both of the drivers. For single-color driving fields, this fully brings into play the nonperturbative physics of HHG, through the intrinsic dipole phase of the harmonics, which is proportional to the field intensity~\cite{lhuillier-calculations-hhg-1992, lewenstein-phase-hhg-1995, zair-quantum-path-2008, hernandez-off-axis-2012}. In that case, an OAM perturbation imprints an azimuthal intensity gradient which, through the intrinsic phase, broadens the OAM content of each harmonic~\cite{rego-nonperturbative-twist-2016}.
Bicircular drivers, on the other hand, allow us to perform a wider exploration of the nonperturbative physics of OAM-HHG, since we can now affect both the field intensity and its shape.

To impart an intensity gradient, as in the single-color case, we divert 10\% of the intensity to a donut-like $\ell=0$ mode (with an amplitude profile ${\sim}r^2e^{-r^2/\sigma^2}$) on both modes; the results are displayed in Fig.~\ref{fig-perturbed-symmetries}(b) and, as in the one-color case~\cite{rego-nonperturbative-twist-2016}, they display an OAM broadening on each harmonic.
To impart a phase gradient, on the other hand, we switch the phase of one of the $\ell=0$ modes, so that the intensity profiles are complementary, giving a constant total intensity but a varying $I_\omega/I_{2\omega}$ intensity ratio. This produces a smaller broadening, shown in Fig.~\ref{fig-perturbed-symmetries}(c), with a different origin: the intensity ratio affects the quantum-path dynamics~\cite{jimenez-control-2018}, which alters both the action and the vector aspects of the recollision~\cite{pisanty-rotating-frame-2017}.
Note that similarly, even in the unperturbed case, quantum-path dynamics and phase-matching effects should contribute to radial structures in the HHG emission~\cite{hernandez-quantum-path-2015, geneaux-radial-LG-2017}.

Finally, it is important to note that this configuration also allows for the production of XUV radiation with controllable OAM, as recently pointed out in Refs.~\citealp{dorney-controlling-2019, paufler-oam-hhg-2018, kong-spin-constrained-2018}. In single-color collinear experiments the OAM scales linearly~\cite{gariepy-creating-2014, rego-nonperturbative-twist-2016, geneaux-synthesis-2016}, with the $q$th harmonic carrying OAM $\ell_q=q\,\ell$~\cite{hernandez-attosecond-vortices-2013, rego-nonperturbative-twist-2016}; for high~$q$, this OAM is often too big to be useful, and it is challenging to detect and characterize in the first place~\cite{gariepy-creating-2014, sanson-hartmann-xuv-wavefronts-2018}. 
One solution is to use a non-collinear perturbing beam~\cite{gariepy-creating-2014, kong-controlling-2017, gauthier-tunable-2017}, but that spreads the harmonic yield over a range of different OAM modes.

Here, however, the use of CR-invariant drivers allows us to imprint the $q^\mathrm{th}$ harmonic with an arbitrary OAM~$\ell_q$ by leveraging the bicircular spin selection rules~\cite{fleischer-spin-2014, pisanty-spin-conservation-2014, alon-selection-rules-1998} to concentrate all of the harmonic yield at a given harmonic order into a single OAM mode. 
Moreover, the interaction region is no longer constrained to lie within the intersection of two non-collinear beams, allowing it to be substantially longer and therefore to support a stronger harmonic emission when properly phase-matched.

The generation of structured high-frequency pulses through TKAM conservation provides a source of bright XUV radiation with customized polarization and OAM. 
For example, it opens a scenario to explore spin-orbit coupling at the nanoscale~\cite{cardano-spin-orbit-photonics-2015}, and to study the modification of photoionization dynamics of single atoms and molecules~\cite{picon-transferring-2010, eckart-ultrafast-2018}. 
Such XUV beams with controlled angular momenta can also be used to enhance light microscopy -- through high-contrast high-resolution spiral-phase imaging~\cite{furhapter-spiral-imaging-2005} or XUV microscopy~\cite{gariepy-creating-2014} --, in lithography~\cite{scott-oam-lithography-2009}, and as tailored waveforms for developing and improving spectroscopic techniques~\cite{rebernik-fel-oam-2017, hernandez-generation-2017}. 
Finally, we note that TKAM conservation in HHG provides a new tool to study ultrafast magnetism, in particular to image magnetic domains~\cite{kfir-nanoscale-2017}, to uncover spin/charge dynamics in magnetic materials~\cite{tengdin-critical-2018} and to generate ultrafast magnetic fields~\cite{blanco-ultraintense-2019}.

\titlespacing*{\section}{0pt}{3mm}{2mm}
\section*{Acknowledgements}
We thank T. Ruchon for helpful observations. 
E.P. acknowledges Cellex-ICFO-MPQ fellowship funding; E.P. and M.L. acknowledge the Spanish Ministry MINECO (National Plan 15 Grant: FISICATEAMO No. FIS2016-79508-P, SEVERO OCHOA No. SEV-2015-0522, FPI), European Social Fund, Fundació Cellex, Generalitat de Catalunya (AGAUR Grant No. 2017 SGR 1341 and CERCA/Program), ERC AdG OSYRIS, EU FETPRO QUIC, and the National Science Centre, Poland-Symfonia Grant No. 2016/20/W/ST4/00314.
A.P. acknowledges funding from Comunidad de Madrid through TALENTO grant ref. 2017-T1/IND-5432.
J.S.R.,  L.P. and C.H-.G acknowledge support from Junta de Castilla y León (SA046U16) and Ministerio de Economía y Competitividad (FIS2013-44174-P, FIS2016-75652-P). 
C.H.-G. acknowledges support from a 2017 Leonardo Grant for Researchers and Cultural Creators, BBVA Foundation and Ministerio de Ciencia, Innovación y Universidades for a Ramón y Cajal contract (RYC-2017-22745), co-funded by the European Social Fund.
L.R. acknowledges support from Ministerio de Educación, Cultura y Deporte (FPU16/02591). We thankfully acknowledge the computer resources at MareNostrum and the technical support provided by Barcelona Supercomputing Center (RES-AECT-2014-2-0085). This research made use of the high-performance computing resources of the Castilla y León Supercomputing Center (SCAYLE, www.scayle.es), financed by the European Regional Development Fund (ERDF).

\bibliographystyle{arthur} 
\bibliography{references}{}


\end{document}